\documentstyle[twocolumn,prl,aps,epsfig]{revtex}

\begin{document}
\draft
\def \beq{\begin{equation}}
\def \eeq{\end{equation}}
\def \beqarr{\begin{eqnarray}}
\def \eeqarr{\end{eqnarray}}

\twocolumn[\hsize\textwidth\columnwidth\hsize\csname @twocolumnfalse\endcsname
\title{Dipolar Excitons, Spontaneous Phase Coherence, 
and Superfluid-Insulator Transition in Bi-layer Quantum Hall Systems 
at $\nu=1$
}

\author{Kun Yang}

\address{
National High Magnetic Field Laboratory and Department of Physics,
Florida State University, Tallahassee, Florida 32306
}

\date{\today}

\maketitle

\begin{abstract}
The spontaneous interlayer phase coherent (111) state of bi-layer Quantum Hall 
system at filling factor $\nu=1$ may be viewed as a condensate of interlayer
particle-hole pairs or excitons. We show in this paper that when the layers are
biased in such a way that these excitons are very dilute, they may be viewed as
point-like bosons. We calculate the exciton dispersion relation, and show that
the exciton-exciton interaction is dominated by the dipole moment they carry.
In addition to the phase coherent state, we also find a Wigner 
Crystal/Glass phase in the presence/absence of disorder which is an insulating
state for the excitons. The position of the phase boundary is estimated and the
properties of the superfluid-insulator type transition between these two phases
is discussed. We also discuss the relation between these ``dipolar" excitons 
and the ``dipolar" composite fermions studied in the context of half-filled 
Landau level.

\end{abstract}

\pacs{73.43.-f,73.21.-b,71.10.Pm,71.35.Lk}
]

Bi-layer quantum Hall systems with spontaneous interlayer phase coherence
(especially at total filling factor $\nu=\nu_1+\nu_2=1$, 
where $\nu_i$ is the filling factor of $i$th 
layer)\cite{wenzee,ezawa,letter,dl1,dl2}
have received renewed interest\cite{balents,stern,fogler,js,demler}, inspired by
a set of recent tunneling measurements in which a spectacular zero-bias peak
was observed at $\nu=1$\cite{jim1}, that resembles the predicted Josephson
effect\cite{wenzee,ezawa}.
Furthermore, the collective mode 
spectrum was mapped out by putting in a parallel component of the external
magnetic field\cite{jim2}, 
and the results are in good qualitative agreement with previous
theoretical calculations\cite{fertig,dl1}. Thus far the theoretical studies have
mostly been focused on the case with equal layer population 
(i.e., $\nu_1=\nu_2
=1/2$), and small interlayer spacing $d$ such that the system is both interlayer
phase coherent and with a quantum Hall charge gap. It was suggested in Ref. 
\onlinecite{demler} on symmetry grounds that other phases, including ones that
are with quantum Hall gap but {\em not} interlayer phase coherent or vice
versa can exist; see also Ref. \onlinecite{kim}. 
These possible new phases, however, were not found in a recent numerical
study\cite{js}, with $\nu_1=\nu_2=1/2$. 

The purpose of this paper is to show by microscopic calculation
that some of these phases can be realized
when the two layers are heavily tilted (by applying a gate voltage), such that
$\nu_1=\delta\nu$ and $\nu_2=1-\delta\nu$, with $\delta\nu\ll 1$. It is well
appreciated that the interlayer coherent state may be viewed as an 
exciton-condensate state in which the particles in one layer and holes in the
other layer form pairs and Bose
condense. The advantage of the limit considered here
is that these excitons are very dilute and their binding energies are much 
higher than their kinetic energies and inter-exciton interactions, thus allowing
these excitons to be treated as point-like bosons. 
Controlled calculations may be
performed on the resultant bosonic Hamiltonian. We find that in addition to the
Bose-Einstein condensed phase for the excitons (that corresponds to the
interlayer-coherent phase), there is also a Wigner crystal/glass phase in the 
presence/absence of disorder, that is
insulating for the excitons. The entire system, on the other hand, exhibits 
quantum Hall effect due to the charge gap. The phase boundary separating these
phases is estimated. The phase transition separating them is the 
superfluid-insulator transition that has been studied extensively in other 
contexts. We also discuss how to detect these phases and the phase transition
experimentally.

{\em The {\rm (111)} State.} We begin by considering the Halperin (111) 
state\cite{halperin} that properly describes the interlayer coherent state:
\begin{equation}
\Psi_{111}(z_i, z_{[k]})=\prod_{i<j}^{N_1}(z_i-z_j)
\prod_{k<l}^{N_2}(z_{[k]}-z_{[l]})\prod_{i, k}(z_i-z_{[k]}),
\label{111}
\end{equation}
where $z_j=x_j+iy_j$ and $z_{[k]}=x_{[k]}+iy_{[k]}$
are the complex coordinates of the
$j$th electron
in layer 1 and $k$th electron in layer 2 respectively, and
$N_1$ and $N_2$ are the numbers of electrons in the
upper and lower layers, which are separately
conserved in the absence of tunneling (assumed in this paper). 
The total number of electrons
$N=N_1+N_2$ equals the Landau level degeneracy.
We have neglected the common exponential factors in the wave function.
The analytic form of the wave function guarantees that it contains no
weight outside the lowest Landau level (LLL).
Unlike most other multicomponent wave functions proposed
by Halperin\cite{halperin}, the $(111)$ wave function is a valid state for
$\nu=1$ for arbitrary relative population in the two layers, an
indication of interlayer coherence and broken symmetry.
In the special case of interlayer distance
$d=0$, states with different $N_1$ are
degenerate. For finite $d$, the two layers are equally populated in the
ground state to minimize Coulomb charging energy in the absence of biasing 
voltage:
$N_1=N_2=N/2$; experimentally 
one can adjust $N_1$ and $N_2$ by applying a finite biasing voltage.
In order to see clearly the pairing
between electrons in one layer and holes in the other, we make a particle-hole
transformation\cite{girvin}
in the lower layer and express the (111) wave function in terms
of coordinates of electrons in layer 1 and {\em holes} in 
layer 2:
\begin{equation}
\Psi_{111}(z_i, \xi_k)\propto {\bf A}\prod_{i=1}^{N_1}e^{z_i\xi_i^*/2}=
{\rm Det}|M_{ij}|,
\label{ph}
\end{equation}
where ${\bf A}$ is an antisymmetrizer, $M_{ij}=e^{z_i\xi_j^*/2}$,
and $\xi_i$ is the complex coordinate of the $i$th hole in
layer 2. Notice that at $\nu=1$, the number of particles in one
layer is always
equal to the number of holes in the other, and that the wave function in the
LLL are analytic functions of $\xi^*$ since holes carry opposite charges.
One can easily show that the states (\ref{111}) and (\ref{ph})
are equivalent by particle-hole transform\cite{girvin} (\ref{ph}) back to
the wave function in terms of coordinates of electrons in both layers.
The state (\ref{ph}) has a simple and clear physical interpretation. It
indicates that every electron in the upper layer is paired up with a hole
in the other layer, with the pairing wave function
\begin{equation}
\Psi_p(z, \xi)=e^{z\xi^*/2}.
\label{pair}
\end{equation}
This is the closest pair that can be formed in the LLL and in fact
represent $\delta(z-\xi)$ projected onto the LLL\cite{gj}.
It has angular momentum zero and hence $S$ wave symmetry.
We thus have shown that the spontaneous interlayer coherence can also be
viewed as $S$ wave
pairing between particles in one layer and holes in the other.

{\em Exciton Dispersion Relation}. Now consider the limit that $N_1=1$ and 
$N_2=N-1$. In this case there is a single particle-hole pair or exciton, and 
Eq. (\ref{pair}) is the {\em exact} ground state wave function of the system
that describes a zero-momentum exciton, which we label as $|0\rangle$. 
The state that describes a single exciton
with finite momentum ${\bf k}$ can be constructed:
\beq
|{\bf k}\rangle = e^{i{\bf k}\cdot {\bf R}}|0\rangle,
\eeq
where ${\bf R}$ is the guiding center coordinate\cite{haldane}
of the electron in layer 1.
The state $|{\bf k}\rangle$ takes the same form as a single spin-wave of a 
spin-polarized filled Landau level\cite{kallin}, 
which is an {\em exact} eigenstate of the
system in both contexts. The $N^2$ allowed ${\bf k}$'s exhaust all the possible
states for a single particle-hole pair. We can calculate the dispersion 
relation of the
single exciton:
\beqarr
E(k)&=&\langle{\bf k}|H|{\bf k}\rangle-\langle 0|H|0\rangle\nonumber\\
&=&{1\over 2\pi}\int_0^{\infty}qV^E(q)e^{-q^2\ell^2/2}[1-J_0(qk\ell^2)]dq,
\eeqarr
where $\ell$ is the magnetic length and $V^E(q)$ is the Fourier transform of
the interlayer Coulomb interaction; neglecting layer thickness we have
$V^E(q)=2\pi e^2 e^{-qd}/(\epsilon q)$,
where $\epsilon$ is the dielectric constant. In the long-wave length limit 
the exciton has a quadratic dispersion:
\beq
E(k)\approx {k^2\over 2m(d)},
\eeq
where the inverse effective mass for the exciton is
\beq
{1\over m(d)}={e^2\ell\over 2\epsilon}\int_0^{\infty}x^2e^{-xd/\ell-x^2/2}dx.
\eeq 
It is worth noting that the origin of the exciton dispersion (or ``kinetic
energy") is solely due to
electron-electron/hole interaction, since the kinetic energy of the electrons
has been quenched
by the strong magnetic field; the momentum of the exciton $k$ is actually a
measure of how closely bound the electron-hole pair is.

\begin{figure}
\vspace{-3truecm}
\centerline{
\epsfxsize=10cm
\epsfbox{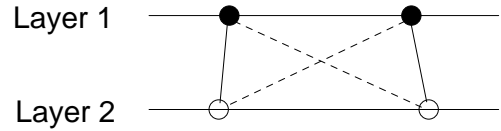}
}
\vspace{-3.2truecm}
\caption{Illustration of two electrons in layer 1 and
two holes in layer 2 forming
two excitons. In principle for the same configuration one may think of two ways
the electrons and holes pair up, as indicated by the solid and dashed lines.
On the other when the pairs are closely bound and separated by large distances,
the solid line description of pairing is clearly more natural and the ambiguity
disappears.
}
\label{fig}
\end{figure}

{\em Exciton-Exciton Interaction}. When $N_1$ is more than 1, we
expect that the electrons in layer 1 form pairs with the exactly same number 
of holes in layer 2 as in the (111) state,
and there will be $N_1$ excitons in the
system. It is very tempting to treat the excitons as point-like bosons and map
the problem onto that of an interacting boson system. 
There is, however, a caveat. When there are more than one excitons, the
naive number of possible states for excitons are bigger than the dimensionality
of the Hilbert space for the electron system. For example, when $N_1=2$, the
dimensionality of the electron
Hilbert space is $N^2(N-1)^2/4$, while when there are 
$N^2$ independent states for a single boson, the number of
states for a pair of bosons would be $N^2(N^2+1)/2$, roughly a factor 
of two bigger. This factor of two can be understood as due to the ambiguity in
the way electrons and holes form pairs, as illustrated in Fig. 1. Thus the
multiple exciton/boson states are overcomplete and obviously 
the overcompleteness problem worsens very quickly as $N_1$ increases.
We believe, however, mapping onto interacting bosons is a valid approach when
$\nu_1=N_1/N\ll 1$ so that the excitons are very dilute, and if we concentrate
only on low-energy properties of the system. This is because at low-energies
the excitons are very closely bound while the inter-exciton distance is very
large; thus the ambiguity illustrated in Fig. 1 disappears because the 
``unnatural" description involves very loosely bound excitons that cost large 
amount energy.

To complete the description of the system in terms of the bosonic excitons, we
calculate the exciton-exciton interaction matrix elements. Consider the
unsymmetrized basis states for two excitons which are normalized:
\beq
|{\bf k}_1{\bf k}_2\rangle=e^{i({\bf k}_1\cdot {\bf R}_1+
{\bf k}_2\cdot {\bf R}_2)}|0\rangle,
\eeq
where $|0\rangle$ here should be understood as the (111) state with particles
1 and 2 in layer 1, and the rest in layer 2. Due to the overcompleteness issue
mentioned above, different basis states are not orthogonal:
\beq
\langle{\bf k}'_1{\bf k}'_2|{\bf k}_1{\bf k}_2\rangle
=-{\delta_{{\bf k}_1+{\bf k}_2,{\bf k}'_1+{\bf k}'_2}
e^{i({\bf k}_1\times
{\bf k}'_1+{\bf k}_2\times {\bf k}'_2)\cdot\hat{z}\ell^2/2}\over N-1}.
\eeq
The Hamiltonian in this case is
\beqarr
H&=&H_0-\Delta H,\\
H_0&=&\sum_{i<j}^N V^A({\bf r}_i-{\bf r}_j),\\
\Delta H&=&\sum_{j>2}^N(\Delta V({\bf r}_1-{\bf r}_j)
+\Delta V({\bf r}_2-{\bf r}_j)].
\eeqarr
Here $V^A$ is the intralayer electron-electron interaction, whose Fourier 
transform is: $V^A(q)=2\pi e^2/\epsilon q$, and $\Delta V
=V^A-V^E$ is the difference between intra- and inter-layer interactions.
We find for small ${\bf k}$'s,
\beqarr
&&\langle{\bf k}'_1{\bf k}'_2|H|{\bf k}_1{\bf k}_2\rangle
={\rm const.}\times\langle{\bf k}'_1{\bf k}'_2|{\bf k}_1{\bf k}_2\rangle
+O(k^2)\nonumber\\
&+&
{2\delta_{{\bf k}_1+{\bf k}_2,{\bf k}'_1+{\bf k}'_2}\over A}
\left[\Delta V_{{\bf k}_1-{\bf k}'_1}-{1\over N}\sum_{\bf q}
\Delta V_qe^{-q^2\ell^2/2}\right], 
\label{inter}
\eeqarr
where $A=2\pi N\ell^2$ is the area of the system. In Eq. (\ref{inter}),
the first term is due to the non-orthogonality of the basis states which has
no physical consequence, and the remaining terms describe the interactions of
the pair of excitons. If we neglect terms of higher order in $k$, we find the
effective interaction between the excitons (in momentum space),
\beq
\tilde{V}_k=2\Delta V_k-{2\over N}\sum_{\bf q}\Delta V_qe^{-q^2\ell^2/2},
\eeq
includes two terms; the first term is a repulsive
dipole-dipole interaction that decays
as $1/R^3$ at large distance $R$; the second term that is $k$-independent 
describes a local attraction that softens the dipole-dipole interaction at
short distances. It can be shown that $\tilde{V}_{k=0} > 0$ so that the overall
interaction is repulsive. The dipole-dipole interaction, of course, comes from
the dipole moment along the $\hat{z}$ direction carried by the exciton due to
charge imbalance between the two layers (see Fig. 1). 

{\em Possible Phases}.
As argued above, for small $\nu_1=N_1/N$ these excitons may be viewed as 
point-like bosons. Such a dilute Bose gas with repulsive interaction supports
two phases. At small $d$, $1/m(d)$ is large while $\tilde{V}$ is small, so the
kinetic energy dominates the physics and the excitons Bose condense; this phase
exhibits spontaneous interlayer phase coherence, and is a
superfluid (SF) phase for the excitons. In particular, in the limit 
$d\rightarrow 0$, $\Delta V$ and $\tilde{V}$ vanish, there is no interaction
among the excitons and they all condense into the zero momentum state,
which is precisely what the (111) state describes; it is the exact ground state
of the system in this limit. The hallmark of this phase is a linear Goldstone
mode whose velocity we estimate using a weak-coupling 
Bogliubov theory\cite{popov}:
\beq
v_g\approx\sqrt{{\nu_1\tilde{V}_{k=0}\over 2\pi\ell^2m(d)}}.
\eeq

On the other hand, for large $d$, $1/m(d)$ is
small and $\tilde{V}$ is large; the repulsive interaction dominates and the
excitons form a Wigner crystal (WC). Upon introducing weak disorder, the WC is
pinned and becomes a Wigner glass; 
thus the WC phase is an insulating phase for the excitons. The fact 
that such a WC phase must exist at large $d$ can also be understood from a 
different viewpoint. Consider the limit $d\rightarrow\infty$ where the two
layers decouple. It is widely expected 
that for small enough $\nu_1$ the 
electrons in layer 1 form a WC\cite{rezayi}. 
The holes in layer 2 will thus form its own WC with 
{\em identical} structure. Thus for large but finite $d$, the two WC's lock
together and electrons and holes pair up; this is exactly the exciton WC. 

It should be emphasized however, that both of these two phases are 
incompressible and the system exhibit $\nu=1$ quantum Hall effect, even though
the {\em neutral} exciton systems are gapless. This is 
because excessive charge in the system will induce unpaired electrons or holes,
which cost a finite amount of energy. Thus the WC phase is an explicit example 
of the phase with quantum Hall effect but no interlayer phase coherence
discussed in Ref. \onlinecite{demler}, in which a broken translational symmetry
was anticipated.

Experimentally, one can easily distinguish between these two phases by 
performing tunneling or drag measurements. In tunneling, one should see a 
zero-bias peak in the SF phase, and Coulomb gap-like behavior in the WC phase;
the Goldstone mode velocity $v_g$ can be measured by applying a parallel 
magnetic field as in Ref. \onlinecite{jim2}. In drag measurement, one should
find that a driving current in one layer always induces the same Hall voltage
in both layers\cite{yang} in the SF phase, while in the WC phase all the current
flow in layer 2 while layer 1 cannot carry any current due to pinning; all 
current going through the sample flow in layer 2 and is carried by the 
background $\nu=1$ electron gas in which the holes are embedded.

{\em Superfluid-Insulator Phase Transition}.
The critical layer spacing separating these two phases may be estimated by 
comparing the (approximate) energies of the corresponding states. For the
(111) state that represents the SF phase, there is no exciton ``kinetic" 
energy and the interaction energy per exciton is $E_{SF}\approx
\nu_1\tilde{V}_{k=0}/(4\pi\ell^2)$; while in the WC phase the 
interaction energy is negligible since the dipolar repulsion vanishes faster
than the kinetic energy in the low density limit; the latter is estimated 
using the uncertainty principle to be 
$E_{WC}\sim \nu_1/(4\pi\ell^2m(d))$.
Comparing the two we obtain a {\em very crude} estimate of the phase boundary
to be $d^*\sim 0.6 \ell$. 

This superfluid-insulator transition is expected to be of 
second-order\cite{fisher1}, at which the conductivity of the {\em excitons},
$\sigma^*_{ex}$, is universal\cite{fisher2}. $\sigma^*_{ex}$ can be
measured in a drag experiment; if the electric fields 
are in the $\hat{x}$ direction in both layers,
then $\sigma^*_{ex}=(J_{1x}-J_{2x})/(E_{1x}-E_{2x})$ at the critical point. 
It should be noted that  
the universality class of this transition is different from the one
in superconductors because i) the excitons are neutral while Cooper pairs are
charged; and ii) due to the absence of time-reversal symmetry, potential 
scattering of the excitons resembles that of scattering off a random 
{\em flux}\cite{green}.

{\em Relation with Dipolar Composite Fermions.} The dipolar excitons discussed
here have close family relations with the dipolar composite fermions
(DCF)\cite{cf,duncan}
studied in the context of half-filled Landau level, and in particular, a 
related model of repulsive {\em bosons} at filling factor 
$\nu=1$\cite{duncan,read}. 
In our case
the exciton is made of an electron in layer 1 and a hole in layer 2; the
latter is a {\em vortex} in the $\nu=1$ background in layer 2. Thus the 
exciton is an electron-vortex pair, just like the DCF.
The fact that the electron and vortex live in separate layers makes the dipolar
nature even more transparent. The fundamental difference is that here the 
exciton is made of two fermionic objects and is therefore a bosonic particle.
In the case of half-filled Landau level the overcompleteness problem similar
to the one discussed here severely limits the usefulness of the
DCFs as elementary particles\cite{duncan}; 
nevertheless it may be useful to study the
low density limit of the DCFs (which is unphysical in that context) as in this
work, where this problem can be avoided.

{\em Acknowledgements}. Some of the key ideas of the present work originated
from (largely unpublished) earlier research on related problems performed in 
collaboration with Duncan Haldane, to whom the author is deeply indebted. He
has also benefited greatly from innumerable discussions with Jim Eisenstein, 
Steve Girvin and Allan MacDonald on this subject. This work was supported by
NSF grant No. DMR-9971541, and the Sloan Foundation.

\end{document}